\documentstyle[12pt,a4wide]{article}
\input epsf

\newcommand{\news}{\setcounter{equation}{0}}
\newcommand{\be}{\begin{equation}}
\newcommand{\ee}{\end{equation}}
\newcommand{\bea}{\begin{eqnarray}}
\newcommand{\eea}{\end{eqnarray}}
\newcommand{\bean}{\begin{eqnarray*}}

\newcommand{\eean}{\end{eqnarray*}}
\font\upright=cmu10 scaled\magstep1
\font\sans=cmss12
\newcommand{\ssf}{\sans}
\newcommand{\stroke}{\vrule height8pt width0.4pt depth-0.1pt}
\newcommand{\Z}{\hbox{\upright\rlap{\ssf Z}\kern 2.7pt {\ssf Z}}}

\newcommand{\C}{{\rlap{\rlap{C}\kern 3.8pt\stroke}\phantom{C}}}
\newcommand{\R}{\hbox{\upright\rlap{I}\kern 1.7pt R}}
\newcommand{\CP}{\C{\upright\rlap{I}\kern 1.5pt P}}
\newcommand{\PP}{\hbox{\upright\rlap{I}\kern 1.5pt P}}

\newcommand{\identity}{{\upright\rlap{1}\kern 2.0pt 1}}

\newcommand{\pibf}{\mbox{\boldmath $\pi$}}
\newcommand{\taubf}{\mbox{\boldmath $\tau$}}

\begin{document}
\pagestyle{plain}
\title{\vskip -70pt
\begin{flushright}
{\normalsize UKC/IMS/96-62} \\
{\normalsize IMPERIAL/TP/95-96/72} \\
\end{flushright}\vskip 50pt
{\bf \Large \bf MULTI-SOLITON DYNAMICS \\ IN THE SKYRME MODEL}\vskip 10pt}

\author{Richard A. Battye$^{\ \dagger}$ and Paul M. Sutcliffe$^{\ \ddagger}$\\[10pt]
{\normalsize $\dagger$ {\sl Theoretical Physics Group, Blackett Laboratory, Imperial College,}}\\{\normalsize {\sl Prince Consort Road, London, SW7 2BZ, UK.}}\\{\normalsize {\sl Email : R.Battye@ic.ac.uk}}\\ 
\\{\normalsize $\ddagger$ {\sl Institute of Mathematics, University of Kent at Canterbury,}}\\
{\normalsize {\sl Canterbury, CT2 7NZ, U.K.}}\\
{\normalsize{\sl Email : P.M.Sutcliffe@ukc.ac.uk}}\\}
\date{}
\maketitle
\vskip 25pt
\begin{abstract}
\noindent We exhibit the dynamical scattering of multi-solitons in the
 Skyrme model for 
configurations with charge two, three and four. First, we construct 
maximally attractive
configurations from a simple profile function and the product ansatz.
 Then using a sophisticated numerical algorithm, initially well-separated
 skyrmions in approximately symmetric configurations 
are shown to scatter through the known minimum energy configurations.  
These scattering events illustrate a number of similarities to BPS monopole configurations of the same 
charge. 
A simple modification of the dynamics to a dissipative regime, allows us to
 compute the minimal energy
 skyrmions for baryon numbers one to four to within a few percent. 
\end{abstract}
\newpage
\section{Introduction}
\news
The Skyrme model was first proposed to describe the strong interactions of
 hadrons in 1962\cite{S}, well before the advent of QCD. In  recent years
 interest in the model has been rekindled with the realization that it is in
 fact a low-energy effective theory for QCD in the large $N_c$ limit\cite
{WIT}. Within the model, topological solitons, known as skyrmions, are 
identified with nucleons and the topological charge with baryon number, $B$.
 Hence, the dynamics of multi-soliton configurations are of considerable 
physical interest since they correspond almost directly to the low-energy 
excitations of nucleons within atoms. This approach has many advantages over
 the the full theory of QCD, the most important of all being that it is 
computational much more accessible to current computer technology than the 
full theory. It is also of considerable mathematical interest due to the 
inherent nonlinear dynamics and the significant numerical problems which 
have to be overcome before realistic comparisons with
the properties of nuclear matter can be made. 

The dynamics of two skyrmions has been studied in previous work using 
analytic met\-hods\cite{AMb,Sch} and also numerical simulations on very small 
grids\cite{AKSS}. It is well understood that two initially well-separated 
skyrmions will scatter at right angles, through the axially symmetric,
 toroidal minimum-energy configuration\cite{V}. This seems to be a geometric
 property of generic two-soliton configurations, for example, BPS 
monopoles\cite{AH} and vortices\cite{She,SheRub}. Here, we investigate the
 dynamics of higher charge configurations up to $B$=$4$ using a sophisticated
 numerical algorithm. We shall see that symmetric configurations of higher
 charge seem to follow a picture similar to that of $B$=$2$, scattering 
--- almost elastically --- through the already known minimum energy 
configurations\cite{BTC,LM}, which have tetrahedral and cubic symmetries for
 $B$=$3$ and $B$=$4$ respectively. Since the collisions are almost elastic,
 very little energy is lost to radiation. Hence, we introduce a phenomenological
 dissipation term which allows us to get to the global minimum energy 
configuration 
without the need for excessive amounts of computer time. The numerical
 methods and higher charge configurations with $B>4$ will be 
discussed in a more detailed publication on this subject\cite{BSutb}. 

The Lagrangian of the Skyrme model may be written in terms
of the SU(2) valued right currents $R_\mu=(\partial_\mu U)U^\dagger$ as
\be
12\pi^2{\cal L} =-\textstyle{1\over 2}
\mbox{Tr}(R_\mu R^\mu)-\textstyle{1\over 16}\mbox{Tr}
([R_\mu,R_\nu][R^\mu,R^\nu])\,.
\label{lag}
\ee
where we have used scaled units of energy and length, 
 and a (+$\,---$) signature for the spacetime metric.
 For the purposes of this letter, we have also ignored the pion mass term,
 since the resulting chiral sigma model represents a more geometrical model
 with a connection to  instantons.

The baryon density ${\cal B}$, whose spatial integral gives the
 integer-valued
 baryon number $B$, is given by
\be
24\pi^2{\cal B}=-\epsilon_{ijk}\mbox{Tr}(R_iR_jR_k)\,,
\ee
where, as throughout this letter, we take latin indices to run over the
spatial values $1,2,3$.
The above units are chosen so that the Fadeev-Bogomolny bound on the 
energy $E$ is simply $E\geq |B|$.

In order to convert to the sigma model notation, we combine
the sigma field and triplet of pion fields by writing 
$U=\sigma+i\pibf\cdot\taubf$
where $\taubf$ denote the Pauli matrices, and the normalization constraint 
which
must be imposed is
\hbox{$\sigma^2+\pibf\cdot\pibf=1$}.  Collecting these fields together in a four 
component
unit vector $\phi=(\sigma,\pi_1,\pi_2,\pi_3)$, the Lagrangian (\ref{lag}) 
becomes 
\be
12\pi^2{\cal L} =\partial_{\mu}
\phi\cdot\partial^{\mu}\phi-\textstyle{1\over 2}
\big{(}\partial_{\mu}\phi\cdot\partial^{\mu}\phi\big{)}^2 + 
\textstyle{1\over 2}\big{(}\partial_{\mu}\phi\cdot\partial_{\nu}
\phi\big{)}\big{(}\partial^{\mu}\phi\cdot\partial^{\nu}\phi\big{)}
+\lambda(\phi\cdot\phi -1)\,,
\ee
where $\lambda$ is the Lagrange multiplier which imposes the chiral 
constraint $\phi\cdot\phi=1$. Using the Euler-Lagrange equations, the field
equations are
\bea
&\big{(}1-\partial_{\mu}{\phi}\cdot\partial^{\mu}{\phi}\big{)}
\Box{\phi}
 -\big{(}\partial^{\nu}{\phi}\cdot\partial_{\mu}\partial_{\nu}
{\phi}-\partial_{\mu}{\phi}\cdot\Box{\phi}\big{)}\partial^{\mu}
{\phi}+\big{(}\partial^{\mu}{\phi}\cdot\partial^{\nu}{\phi}\big{)}
\partial_{\mu}\partial_{\nu}{\phi}\cr
& +
\big{[}\big{(}\partial_{\mu}{\phi}\cdot\partial_{\nu}{\bf\phi}\big{)}
\big{(}\partial^{\mu}{\phi}\cdot\partial^{\nu}{\phi}\big{)}
+(1-\partial_{\mu}{\phi}\cdot\partial^{\mu}{\phi}\big{)}
\partial_{\nu}{\phi}\cdot\partial^{\nu}{\phi}
\big{]}\phi
=0\,,\label{eom}
\eea
where $\Box$ denotes the wave operator in $(3+1)$-dimensions.

This equation has a number of numerical problems associated with it\cite{CB}.
The first is a standard instability of the model itself. When the kinetic 
energy is locally greater than the potential energy the assumptions under
 which the model is deduced from QCD are invalidated and the equation becomes
 elliptic, rather than hyperbolic. Therefore any numerical method designed to
 solve the equations in the hyperbolic regime will become unstable in the
 elliptic regime and vice-versa. Since this is an instability of the model, rather than one of any numerical scheme which may be employed, 
we will be restricted to studying configurations which do not have large 
amounts of kinetic energy. In particular, we will not be able to investigate
 those Lorentz boosted to ultra-relativistic velocities and those which emit large
 amounts of radiation. 

The other more numerical problems are related to the overall Courant 
stability and also the imposition of the chiral constraint. The precise resolution of these problems will be 
discussed in later work, but it suffices to say that it is possible to 
construct an algorithm which is stable for many time steps ($>1000$) with 
values of the spatial step size $\Delta x$ and the time step $\Delta t$ 
within half an order of magnitude of the standard Courant criterion 
$\Delta x=\sqrt{3}\Delta t$. The simulations presented in this letter used
 fourth order spatial differences on grids which range in size from $70^3$
 points to $100^3$ points, depending upon the particular simulation, with 
$\Delta x=0.1$ and the range
$0.01\le \Delta t\le 0.02.$ 

The field of a static single skyrmion centred at the origin has the hedgehog
 form,
\be
U({\bf x})=\mbox{exp}[if(r)\widehat{\bf x}\cdot\taubf]\,,
\ee
where $r=\vert {\bf x}\vert$ and the profile function $f(r)$ has boundary
conditions $f(0)=\pi$ and $f(\infty)=0$. Solving numerically for this
profile function\cite{ANW} determines the energy of a single skyrmion to be
$E=1.232$. The centre of the skyrmion can be placed at any point in space
by a translation, and its isospin orientation can be changed by conjugation 
of the
$U$ field by any fixed element of $SU(2)$. It can also be made to move
with a fixed velocity by Lorentz boosting the static solution.

Instead of using the numerical profile function, we shall use a simple analytic 
approximation based on sine-Gordon kinks\cite{Sa},
\be 
f(r)=4\mbox{tan}^{-1}(e^{-r})\,,\label{kink}
\ee
which gives a skyrmion with energy less than $1\%$ above the true skyrmion
 energy.
Since we are using this configuration only for our initial conditions,
it will be an adequate approximation. Note that (\ref{kink}) 
has an exponential decay, whereas in the massless pion limit the true profile
function has a power-law decay. An approximate profile function with the
 correct
decay behaviour can be obtained by computing the holonomy of 
instantons\cite{AMa}.
However, this approximation has slightly more energy than  (\ref{kink}), 
which
implies that the kink approximation is better in the region of interest near the core of the
skyrmion. 

By taking a product ansatz $U=U_1U_2$ of two well-separated single skyrmion 
fields $U_1$ and $U_2$, a $B$=$2$ configuration can be constructed which 
models 
two well-separated and undistorted skyrmions. The interaction between the 
two skyrmions can be repulsive or attractive depending upon the relative
 isospin
orientation of the two skyrmions\cite{JJP}. This attraction is
 maximal if one skyrmion is rotated relative to the other through an angle of $180^\circ$ about an axis orthogonal to the line joining the two skyrmions. Two skyrmions
which have such a relative orientation are said to be in the attractive
 channel.

\section{Multi-soliton dynamics}
\news
The simplest scattering event involves the head-on collision of two skyrmions
in the attractive channel. As in all our simulations we begin with a configuration
of well-separated skyrmions obtained from the product ansatz.
The precise configuration used consists of two skyrmions with positions
\be
{\bf X}_1=(0,0,a),\  {\bf X}_2=(0,0,-a)\
\ee
where  $a$=$1.5$, the relative orientation given by a $180^\circ$ rotation
around the $y$-axis and each skyrmion boosted to a velocity of $v$=$0.3$. 
Fig.~1 shows an isosurface plot of baryon density ${\cal B}$ at regular 
intervals\footnote{We display all our results in the form of isosurfaces of 
baryon density, but the isosurfaces of constant energy density are almost
 indistinguishable.}. As the initially well-separated skyrmions come together
 they distort, eventually reaching the expected torus followed by $90^\circ$
 scattering.
 The interaction is almost elastic with very little
 radiation involved in the scattering process.
The skyrmions do, of course, attract once more and pass through the torus 
again
with a $90^\circ$ scattering. This process repeats, with multiple
right angle scatterings and the amplitude of oscillation around the torus 
decreasing with time. 
At first sight it may seem surprising that so little radiation is observed
in the skyrmion scattering event, particularly since we are considering
the model without a pion mass term. However, from studies of 
baby skyrmions\cite{PSZ}, it appears that the production
of radiation is counter-intuitive; less radiation is observed in the case
where it is massless. Superficially at least, our results appear to support these findings.

For the case of charge three, we choose an initial configuration which will
 scatter close to the minimal energy tetrahedral
$B$=$3$ skyrmion. It turns out that in choosing skyrmion initial
 configurations
it is a good guide to first consider the analogous case in the context of BPS monopoles.
It is known that a tetrahedral 3-monopole exists\cite{HMM,HSa}, which is
 formed
during the scattering of three monopoles with cyclic $C_3$
 symmetry\cite{HMM,Sb}.
The monopoles are initially well-separated on the vertices of a large
 contracting
equilateral triangle.
 We therefore take three well-separated skyrmions in such
a configuration. In the skyrmion case, we must also choose the initial 
orientations. We do this so as to maximize the amount of attraction,
which in this case can be done by choosing each pair of skyrmions to be in the 
attractive channel. 
Take the positions of the skyrmions to be ${\bf X}_i$, and
define the relative position vectors ${\bf X}_{ij}={\bf X}_i-{\bf X}_j$.
Let ${\bf n}_{ij}$ be the unit vector such that the relative orientation
 between the
skyrmion at ${\bf X}_i$ and the one at ${\bf X}_j$ is given by a rotation
by $180^\circ$ around the axis ${\bf n}_{ij}$.  
Explicitly, we take
\bea
&{\bf X}_1=(-a,-a,-a), \ 
{\bf X}_2=(-a,a,a), \
{\bf X}_3=(a,-a,a), \cr
& {\bf n}_{12}=(1,0,0), \ {\bf n}_{13}=(0,1,0).\label{tri}
\eea
This implies that ${\bf n}_{23}=(0,0,1)$ and it is easily checked that this
choice satisfies the requirement of all pairs being in the attractive channel ie.
${\bf X}_{ij}\cdot {\bf n}_{ij}=0$  for all $i\neq j$.

Again, we choose $a=1.5$, but this time each skyrmion is boosted to have an initial velocity of $v=0.1\sqrt{3}$ towards the centre of the triangle. 
The evolution of this configuration is shown in fig.~2. We should point out
 that, although we have constructed a configuration
with cyclic symmetry,
this symmetry is broken by the product ansatz implementation $U=U_1U_2U_3$,
 which is clearly asymmetric under permutations of the indices.

We see that initially there are three separated skyrmions on the vertices of
an equilateral triangle, which deform as they coalesce. But each
skyrmion behaves slightly differently, due to the symmetry breaking product
 ansatz.
Clearly the greater the initial separation of the skyrmions then the closer 
the
product ansatz configuration is to having cyclic symmetry. The dynamics is,
nonetheless, remarkably similar to the monopole case\cite{Sb}, except for the influence of the potential and the approximate nature of the symmetry. The skyrmions form
an approximately tetrahedral configuration, which then splits into a single
 skyrmion
and a charge two torus. This may well be one of the important vibrational
 modes of the tetrahedral skyrmion which needs to be considered in a 
quantization of the
classical solution. 

In the monopole case, a second scattering process through the tetrahedral
 3-monopole
is also known\cite{HSc}, which involves a twisted line scattering of three
collinear monopoles. We now investigate whether a similar scattering process
can occur for three skyrmions by
taking the positions of the three collinear skyrmions to be
\be
{\bf X}_1=(0,0,a), \ 
{\bf X}_2=(0,0,0), \
{\bf X}_3=(0,0,-a).
\ee
To put the first and second skyrmions in the attractive
channel we may, without loss of generality, take 
${\bf n}_{12}=(1,0,0)$. For the second and third
skyrmions to be in the attractive channel requires that the
axis ${\bf n}_{23}$ be in the \hbox{$x_1x_2$-plane.} Furthermore,
if the relative orientation between the first and third
skyrmions is to be given by a $180^\circ$ rotation about
some axis then we must have that ${\bf n}_{12}\cdot {\bf
n}_{23}=0$, which determines that ${\bf n}_{23}=(0,1,0)$.
Thus, we deduce that ${\bf n}_{13}=(0,0,1)$,
but ${\bf X}_{13}\cdot {\bf n}_{13}
\ne 0$. So the first and third skyrmions are not in the
attractive channel; in fact, they are in a repulsive channel.
However, one may argue
that the interaction between the first and third skyrmion
should not be classified in this naive
way, since the second skyrmion lies directly between the
two and thus distorts the dipole interaction.

We therefore proceed with the above configuration, which incidentally
has the twisted line symmetry (up to isospin rotations)
of inversion plus $90^\circ$ rotation, with the value 
$a$=$1.5$ and the two outer skyrmions boosted towards the 
one at the origin with a velocity $v$=$0.1$. The results are displayed in fig.~3. The initial
configuration is  interesting, since it does not resemble three
well-separated skyrmions even though we used the product ansatz.
If we  increase the initial separation $a$,
then three well-separated skyrmions do appear. However, at
this separation we obtain a twisted figure-of-eight shape
which is remarkably similar to the corresponding monopole
configuration\cite{HSc}. Once again, the similarity to monopole
dynamics is quite striking: an approximate
tetrahedral skyrmion is formed, before the motion continues
towards a toroidal configuration.

It is known that a toroidal $B$=$3$ skyrmion has a relatively high energy,
and with these initial conditions the combined energy of
the skyrmions is not sufficient to carry it through to the
torus. Therefore, it returns back through similar dynamics, up to
a slight rotation. This contrasts sharply with the monopole case
where all configurations have equal energy and the dynamics
continues through the torus and forms the dual tetrahedron
\cite{HSc}. 
In the skyrmion case, the dynamics is influenced by the potential which prevents
 the formation of the toroidal configuration. This is an important difference between 
skyrmions and BPS monopoles.
The above type of skyrmion scattering event has previously
been conjectured in ref.\cite{W} based upon JNR instanton
holonomy calculations. However, it was not possible to obtain
configurations of well-separated skyrmions due to the 
restriction to JNR instantons rather than the general instanton.

Four monopoles on the vertices of a contracting regular tetrahedron scatter
through a cubic monopole\cite{HSa}. Therefore, we now consider an analogous four skyrmion
scattering process. The starting point is the $B$=$3$ system given by 
(\ref{tri}), where the skyrmions may be considered as positioned on three
vertices of a regular tetrahedron. To this system we add a fourth skyrmion placed
at the remaining vertex ${\bf X}_4=(a,a,-a)$ with orientation given by
${\bf n}_{14}=(0,0,1)$. The additional relative orientations are then determined
to be
${\bf n}_{24}=(0,1,0)$ and ${\bf n}_{34}=(1,0,0)$ so that again we
have ${\bf X}_{ij}\cdot {\bf n}_{ij}=0$  for all $i\neq j$, showing that
all pairs of skyrmions are in the attractive channel. Again, we use $a$=$1.5$, but this time with no initial Lorentz boosts.

The evolution of this configuration is displayed in fig.~4. The mutual attractions cause the skyrmions to coalesce
and form a cubic configuration, before scattering out to lie on the vertices of
a tetrahedron dual to the initial one. Again the product ansatz implementation
slightly distorts the above description, resulting in the symmetries being
only approximately attained. Up to this technicality, however, the scattering event is
once again a close copy of the monopole case\cite{HSa}. Note that a one-parameter
family of ADHM instanton generated skyrmions exists which is a good approximate
description of this process\cite{LM}. By taking an initial configuration
from this family, rather than using the product ansatz, the above approximate
symmetries could be made exact.

\section{Minimum energy configurations}
\news

If we were able to run the simulation for long enough, 
each of the attractive configurations would eventually settle down to the
 global minimal energy configuration. However, given the small amounts of
 radiation emitted in a single scattering event this is not computationally feasible. Instead, we modify
the dynamics to a dissipative regime by the inclusion of a friction term
in the equation of motion. That is, we add a phenomenological term
 $\epsilon\dot\phi$, with $\epsilon=0.5$, to the equations of motion
 (\ref{eom}). This does not affect the static solutions of the model, but enables the
 configuration to settle to a static solution in a much shorter period of
 time.  Of course there is no guarantee that this static solution will be the
 minimum energy configuration, but given the lack of symmetry imposed by the
 product ansatz it is likely to be. In fig.~5 we plot a ${\cal B}$
surface for the minimal energy skyrmions of charge one to four and we
 also compute their energy and baryon number, which are given in Table 1. 

Numerical effects, particularly the fact that we work on a grid covering only
 a
finite volume of space, mean that there will be some inaccuracies in
 calculating both $B$ and $E$. For example, in the charge two case we obtain 
$B=1.971$ rather than the integer value 2. However, our results are more
 accurate than the previous numerical study of multi-skyrmions\cite{BTC}, 
with all our calculations of the baryon number being considerably closer to
 integer values. Moreover, our computations are the
first to be performed for the massless pion case, which makes the
calculations more difficult since the fields are only localized power-like 
rather than exponential. By calculating the energy per baryon $E/B$ we should
 reduce the severity of these inaccuracies.
 The effects of the lost tails will not cancel
out exactly --- the baryon density contains three derivatives and the two terms
in the energy density contain two and four derivatives --- but it should correct
the results in the right direction. For comparison we use our scheme to compute
these quantities for the single skyrmion, where the result is accurately known
to be $E/B=1.232$.  

It is possible to obtain approximate skyrmions by computing the holonomies
of inst\-antons\cite{AMa}. This provides a method of computing upper bounds
 for
the energies of the minimal energy skyrmions of each baryon number.
Results are available for $B=1,2,3,4$ and the Skyrme crystal 
\cite{AMa,AMb,LM,MS}. For comparison we include in Table 1 the instanton 
computed values for $E/B$ which we denote by $E_I$. These results seem to overestimate the energy by around $1\%$ and in the case of 
$B$=$1$ are significantly less accurate than our calculations. We suggest
that this is likely to be true for the higher charge configurations.

\begin{center}
\begin{tabular}{|c|c|c|c|c|} \hline

charge &  $B$ & $E$ & $E/B$ & $E_I$\\
\hline
1 & 0.984 & 1.214 & 1.233 & 1.243\\
2 & 1.971 & 2.309 & 1.171 & 1.192\\
3 & 2.965 & 3.407 & 1.149 & 1.163\\
4 & 3.945 & 4.378 & 1.110 & 1.132\\
\hline
\end{tabular}
\end{center}
{\bf Table 1 } : Calculated values of the total energy for charge one to 
four in the massless pion limit. For comparison to previous work, use $E/B$.

\section{Conclusions}
\news
We have performed numerical simulations of skyrmion dynamics, with initial
conditions created from a product ansatz of single skyrmions, 
using state of the art computer technology. By considering
symmetric configurations, motivated by BPS monopoles,
we have displayed scattering processes which form the minimal energy
configurations for baryon numbers two to four. 
However, the analogy to monopoles is not exact, since in the skyrmion case
 the dynamics is influenced by the potential.
Nonetheless, it seems possible to construct similar scattering events
associated with platonic solids already seen in the monopole
case.

By inclusion of a dissipation term, we were able to accurately (within 1\%)
 compute the energies of the known bound states. We see that the cubic $B$=$4$
configuration is the most tightly bound, reflecting the behaviour of 
low-energy baryons. This makes creating higher charge configurations from the 
product ansatz more difficult, since, for example, a general $B$=$5$ 
configuration has tendency to split up into a cubic $B$=$4$ configuration and 
a single skyrmion.
In fact, it is not possible to arrange more than four well-separated skyrmions
so that all pairs of interactions are in the attractive channel.
However, by going beyond a simple product ansatz approach we have been
able to construct skyrmion bound states for baryon numbers greater than four.
These results, together with other saddle point configurations and
scattering events, will be presented in an future publication\cite{BSutb}.

\section*{Acknowledgements}
\news

RAB acknowledges the support of PPARC Postdoctoral fellowship grant GR/K94799 
and PMS thanks the Nuffield Foundation for a newly appointed lecturer award. 
We also acknowledge the use of the SGI Power Challenge at DAMTP in Cambridge 
supported by the PPARC Cambridge Relativity rolling grant and EPSRC Appled 
Mathematics Initiative grant GR/K50641. Thanks to Paul Shellard for his 
understanding in these respects and to Brad Baxter for discussions on the
numerical aspects.

\def\jnl#1#2#3#4#5#6{\hang{#1 [#2], {\it #4\/} {\bf #5}, #6.}}
\def\jnltwo#1#2#3#4#5#6#7#8{\hang{#1 [#2], {\it #4\/} {\bf #5}, #6;{\bf #7} #8.}}
\def\prep#1#2#3#4{\hang{#1 [#2],`#3', #4.}} 
\def\proc#1#2#3#4#5#6{{#1 [#2], in {\it #4\/}, #5, eds.\ (#6).}}
\def\book#1#2#3#4{\hang{#1 [#2], {\it #3\/} (#4).}}
\def\jnlerr#1#2#3#4#5#6#7#8{\hang{#1 [#2], {\it #4\/} {\bf #5}, #6.
{Erratum:} {\it #4\/} {\bf #7}, #8.}}
\def\prl{Phys.\ Rev.\ Lett.}
\def\pr{Phys.\ Rev.}
\def\pl{Phys.\ Lett.}
\def\np{Nucl.\ Phys.}
\def\prp{Phys.\ Rep.}
\def\rmp{Rev.\ Mod.\ Phys.}
\def\cmp{Comm.\ Math.\ Phys.}
\def\mpl{Mod.\ Phys.\ Lett.}
\def\apj{Ap.\ J.}
\def\apjl{Ap.\ J.\ Lett.}
\def\aap{Astron.\ Ap.}
\def\cqg{Class.\ Quant.\ Grav.} 
\def\grg{Gen.\ Rel.\ Grav.}
\def\mn{M.$\,$N.$\,$R.$\,$A.$\,$S.}
\def\ptp{Prog.\ Theor.\ Phys.}
\def\jetp{Sov.\ Phys.\ JETP}
\def\jetpl{JETP Lett.}
\def\jmp{J.\ Math.\ Phys.}
\def\zpc{Z.\ Phys.\ C}
\def\cupress{Cambridge University Press}
\def\pup{Princeton University Press}
\def\wss{World Scientific, Singapore}

\section*{Figure Captions}  

\noindent Fig.~1. Head-on collision of two skyrmions in the attractive channel 
illustrating the right-angled scattering.
\bigskip

\noindent Fig.~2. Three skyrmion scattering with cyclic symmetry to form 
tetrahedron.
\bigskip

\noindent Fig.~3. Twisted line scattering of three skyrmions 
through a
 tetrahedron. The configuration does not reach the charge 
three torus due 
to the potential, illustrating the difference between skyrmions
 and BPS 
monopoles.
\bigskip

\noindent Fig.~4. Four skyrmion with tetrahedral symmetry
 scattering through 
a configuration with approximate cubic symmetry.
\bigskip

\noindent Fig.~5. Minimal energy skyrmions for 
baryon numbers one to four.\\

\noindent {\sl Note: Figures are attached as gif files. Type
 {\bf xv sky1fig1.gif}\ etc. to view figures. Hard copies
can be obtained on request by emailing p.m.sutcliffe@ukc.ac.uk}

\end{document}